\documentclass[twocolumn,reprint,amsmath,amssymb,prx]{revtex4-1}
\usepackage{graphicx}
\usepackage{dcolumn}
\usepackage{textcomp}
\usepackage{amssymb}
\usepackage{amsmath}
\usepackage{color}
\usepackage{footnote}
\usepackage{microtype}
\usepackage{bm}
\usepackage{changes}
\usepackage{hyperref}

\makeatletter
\newcommand*{\rom}[1]{\expandafter\@slowromancap\romannumeral #1@}
\makeatother

\newcommand{\beq}{\begin{equation}}
\newcommand{\eeq}{\end{equation}}
\newcommand{\beqa}{\begin{eqnarray}}
\newcommand{\eeqa}{\end{eqnarray}}

\begin{document}
\title{Ferroelectric control of charge-to-spin conversion in WTe$_{2}$}
\author{Homayoun Jafari, Arunesh Roy, and Jagoda S\l awi\'{n}ska}%
\affiliation{Zernike Institute for Advanced Materials, University of Groningen, Nijenborgh 4, 9747 AG Groningen, The Netherlands}

\date{\today}

\begin{abstract}
Ferroelectric materials hold great potential for alternative memories and computing, but several challenges need to be overcome before bringing the ideas to applications. In this context, the recently discovered link between electric polarization and spin textures in some classes of ferroelectrics expands the perspectives of the design of devices that could simultaneously benefit from ferroelectric and spintronic properties. Here, we explore the concept of non-volatile ferroelectric control of charge-to-spin conversion in semimetallic WTe$_2$, which may provide a way for non-destructive read-out of the polar state. Based on the first-principles simulations, we show that the Rashba-Edelstein effect (REE) that converts electric currents into spin accumulation switches its sign upon the reversal of the electric polarization. The numerical values of REE, calculated for the first time for both bulk and bilayer WTe$_2$, demonstrate that the conversion is sizable, and may remain large even at room temperature. The ferroelectric control of spin transport in non-magnetic materials provides functionalities similar to multiferroics and allows the design of memories or logic-in-memory devices that combine ferroelectric writing of information at low power with the spin-orbit read-out of state.
\end{abstract}
\maketitle


\textit{Introduction.} Our growing needs to store and process information require alternative approaches to make electronics faster and more power-saving \cite{beyondvoneumann}. Ferroelectric (FE) materials are exciting candidates for design of novel computing architectures, offering non-volatile, high-speed writing at low voltage along with high endurance, but challenges related to storage density and detection of state limits their use in memories and logic applications \cite{review_memories, review_2D_ferro}. In this context, a recent concept of a magneto-electric spin-orbit (MESO) device proposed by Manipatruni \textit{et al.} may become a game-changer for electronic devices \cite{manipatruni2019Meso}. It employs the ferroelectric state of a multiferroic material to perform logic operations. At the same time, the read-out can be efficiently accomplished via the conversion between charge and spin currents based on the interface with topological insulators. Compared with standard complementary metal-oxide-semiconductor (CMOS) technology, the prototype stands out for much lower switching voltage of only 100 mV, increased logic density, and non-volatility that allows for very low standby power. Nevertheless, it contains various quantum materials, including multiferroics, which makes it difficult to realize and control.

In parallel, progress in spintronics brought novel routes to harness and control spins via different stimuli. The mechanisms of charge-to-spin conversion (CSC), such as spin Hall effect (SHE) \cite{she_rev} and Rashba-Edelstein effect (REE) \cite{edelstein1990spin,Ivchenkov1978photogalvanic,Aronov1990mesoscopic} found in various conventional and topological materials, rely on spin-orbit coupling (SOC) of electronic states to generate spin accumulation in response to an applied charge current. Moreover, recent studies revealed that changing the sign of electric polarization may reverse CSC, which provides an alternative way of polar state detection. FE writing could be then combined with the spin-orbit (SO) read-out within one material without additional interfaces. The ferroelectric control of CSC was observed in the two-dimensional electron gas (2DEG) at the ferroelectric-like surface of SrTiO$_{3}$, where REE originates from the presence of spin-polarized electronic states \cite{bibesNatSTO}. The switching of ferroelectric polarization reverses the spin texture and thus, the sign of generated spin accumulation. Although similar ferroelectric manipulation of spin currents was successfully realized in the thin film of GeTe \cite{varotto2021Natelec}, a vaster choice of CSC materials that are ferroelectric at room temperature is still desired.

In this Letter, we explore the possibility of non-volatile ferroelectric control of CSC in orthorhombic WTe$_2$, a recently revealed polar semimetal. The ferroelectricity of two- and three-layer WTe$_2$ was linked to its layered nature and ultrathin thickness \cite{fei, tunable_semimetal, berry_memory, ren}; yet the spontaneous polarization was confirmed also in its bulk phase at room temperature \cite{tsymbal_scienadv}. At the same time, CSC in WTe$_2$ was separately reported in experimental studies \cite{marcos_nature, macneill, saroj}. The relatively strong SOC may lead to both SHE and REE, whereby the contributing states are located around the Fermi level, making CSC easily accessible even without doping. Here, we use the recently implemented approach based on density functional theory (DFT) to numerically estimate REE in the bilayer and bulk WTe$_2$ at different temperatures and demonstrate that it reverses for opposite ferroelectric ground states, in addition to large conversion efficiency, which may persist up to room temperature. While both bilayer and bulk WTe$_2$ show high potential for novel electronic devices, including possible logic-in-memory applications, the spin accumulation in these two systems may allow different functionalities.

\textit{Computational method.} Our first-principles density functional theory (DFT) calculations were carried out using the Quantum Espresso package \cite{giannozzi2009quantum, qe2}. We used the Perdew-Burke-Ernzerhof (PBE) exchange-correlation functional \cite{pbe}, along with the fully relativistic projector-augmented wave (PAW) potentials and PBE-D3 method to account for the van der Waals interactions \cite{dft-d3}. The kinetic energy cutoff for the plane wave basis was set to 110 Ry. The unit cells were constructed using the experimental lattice constants equal to $a$=3.47 \AA{}, $b$ = 6.24 \AA{}, $c$ = 14.02 \AA{}, and $a$=3.48 \AA{}, $b$ = 6.27 \AA{} for bulk \cite{mar1992metal} and bilayer \cite{bi_lattice_param}, respectively. In case of the bilayer, we employed the slab geometry with the vacuum layer of 17 \AA, in order to avoid interaction between spurious replicas. The internal atomic positions were relaxed until the energies were converged within $10^{-8}$ eV and the maximum forces on atoms were below 10$^{-3}$ eV$\cdot$\AA$^{-1}$. The Brillouin zones (BZ) were sampled on the Monkhorst–Pack $k$-point grids of 16$\times$10$\times$4 and 16$\times$10$\times$1. The post-processing calculations of REE were performed in the framework of semi-classical Boltzmann transport theory using the \textsc{paoflow} code \cite{paoflow1, paoflow2}, following the details given elsewhere \cite{chiral_tellurium}. The ultra-dense $k$-grids of $216 \times 144 \times 6$ and $176 \times 96 \times 1$ were employed to interpolate the Hamiltonian for the accurate convergence of REE. For $T=0$ the derivative of the Fermi distribution is equal to $\delta(E_\mathbf{k}-E)$ and we approximated it with a Gaussian function, while for $T>0$ we directly calculated the derivative. The influence of the temperature on the electronic structure itself was not taken into account.

\begin{figure}[t!]
		\includegraphics[scale=1.0]{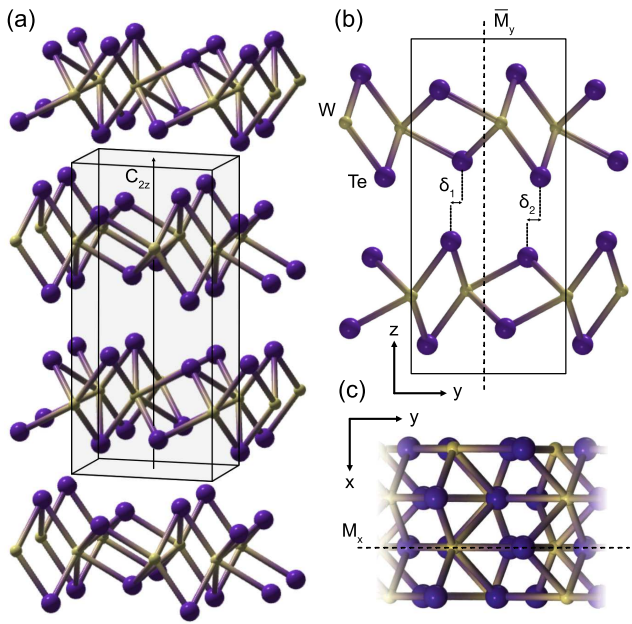}
	\caption{Structure of bulk WTe$_2$ with the polarization upwards ($P_{\uparrow}$). (a) Perspective view. (b) Side view. (c) Top view. The displacements $\delta_1 = 0.23$ and $\delta_2 = 0.30$ \AA, describe the ferroelectric distortion of the state with $P_{\uparrow}$. The configuration $P_{\downarrow}$ (not shown) can be obtained by sliding the layer so that $\delta_1 \rightarrow -\delta_2$ and $\delta_2 \rightarrow -\delta_1$ \cite{ren}. Dashed lines denote the symmetries, the mirror reflection $M_x$, the glide reflection $\overline{M}_y$ and the two-fold screw rotation $\overline{C}_{2z}$.}
	\label{fig:1}
\end{figure}


\textit{Structure and ferroelectricity.} The bulk WTe$_{2}$ crystalizes in $T_d$ phase described by the space group (SG) \#31 ($Pmn2_1$) and the point group $C_{2v}$. The orthorhombic unit cell shown in Fig. 1(a) contains 12 atoms arranged in two layers weakly interacting via van der Waals forces, while each layer consists of W atoms strongly bonded to Te atoms above and below, and forming a zigzag chain along the $y$ direction. The crystal is invariant with respect to four symmetry operations: (1) the identity $I$; (2) the mirror reflection $M_x$, (3) the glide reflection $\overline{M}_y$ consisting of a mirror reflection with respect to $y$ followed by a fractional translation by a vector $\tau$ = $(0.5a, 0, 0.5c)$, and (4) the two-fold screw rotation $\overline{C}_{2z}$ combining two-fold rotation around the $z$-axis with the translation by $\tau$. The bilayer WTe$_2$ has essentially the same structure, but the periodicity is broken along the $z$ axis. The space group is thus reduced to \#6 ($Pm$) and the only symmetries are the identity $I$ and the mirror reflection $M_x$.

While the origin of the ferroelectricity in WTe$_2$ is still debated \cite{fei}, the electric polarization ($\vec{P}$) is usually assigned to the tiny distortion of Te atoms denoted in Fig. 1(b) as $\delta_1$ and $\delta_2$. These in-plane displacements cause charge transfer in the region between the layers, which in turn induces the electric dipole along the $z$ axis. Thus, the out-of-plane polarization is related to the in-plane displacement \cite{physchemlett}, and the transition between the $P_{\uparrow}$ and $P_{\downarrow}$ states can be achieved by sliding the layers with respect to each other by the distance $\delta_1+\delta_2$. The calculated distortions for bulk, $\delta_1 = 0.23$ and $\delta_2 = 0.30$ \AA, are very similar to the values that we obtained for the bilayer, $\delta_1 = 0.22$ and $\delta_2 = 0.28$ \AA; we note that the latter perfectly agree with the previously reported values \cite{ren}. Surprisingly, the electric polarization for the bulk and bilayer differ by an order of magnitude. The ionic contribution that we estimated following the approach from Ref. \cite{tsymbal_scienadv} equals to $\vec{P}_{\mathrm{ion}}$=1.0 $\mu $C/cm$^2$. In case of the bilayer, the calculated equivalent of the volume polarization is $\vec{P}_{\mathrm{ion}}=0.037$ $\mu $C/cm$^2$ which agrees with the experimentally reported magnitude (0.06 $\mu $C/cm$^2$) \cite{tunable_semimetal}. We note that these values are relatively small compared with conventional ferroelectrics (e.g. 20 $\mu $C/cm$^2$ for BaTiO$_3$).

\begin{figure*}[ht]
		\includegraphics[scale=0.44]{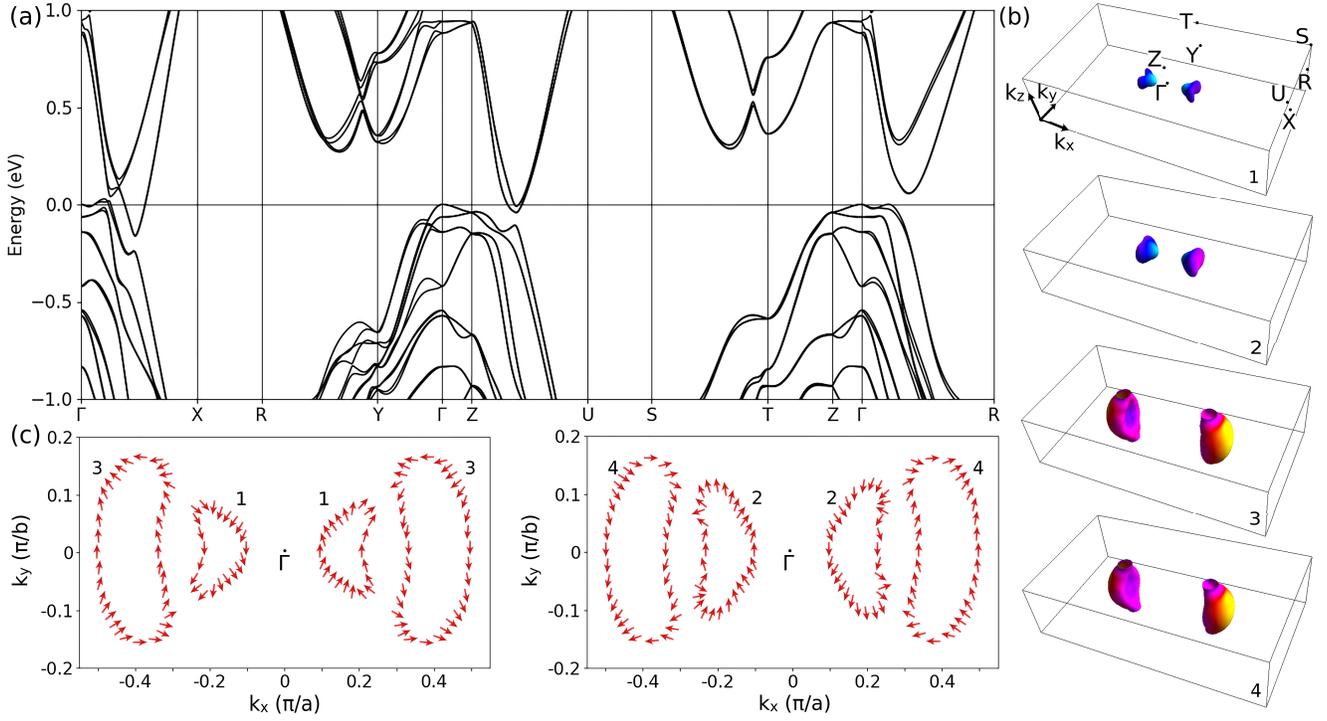}
	\caption{Electronic properties of bulk WTe$_2$ with polarization upwards ($P_{\uparrow}$). (a) Band structure calculated between the high-symmetry points defined in (b). (b) Fermi surface consisting of four bands (1-4) shown in separate panels. The color scheme reflects the local band velocity \cite{FermiSurfer}. (c) Constant energy contours ($E=E_F$) plotted at the $k_x-k_y$ plane for $k_z=0$. The arrows correspond to (S$_{x}$, S$_{y}$) components of the spin texture along the Fermi contours, indicating the spin direction. The $S_z$ component of the spin texture is negligible and it is not shown. The numbers correspond to Fermi sheets displayed in (b).}
	\label{fig:2}
\end{figure*}

\begin{figure*}[ht]
		\includegraphics[width = 0.99\textwidth]{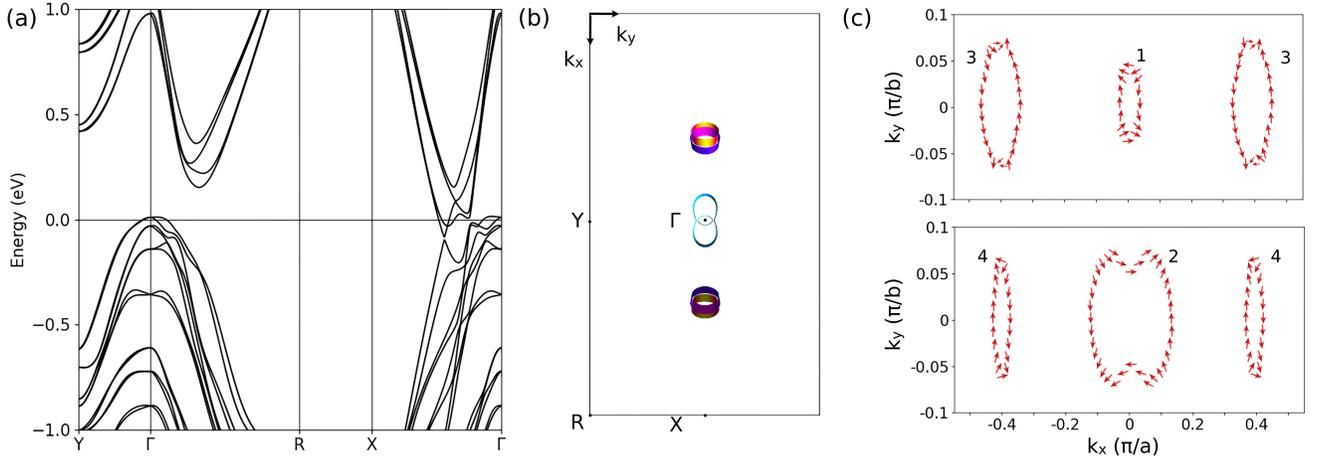}
	\caption{Electronic properties of bilayer WTe$_{2}$ with polarization upwards ($P_{\uparrow}$). (a) Band structure calculated along the high-symmetry lines in the BZ; the high-symmetry points are displayed in (b). (b) Fermi surface ($E=0$) consisting of four bands; the full BZ is displayed for convenience. (c) Spin texture along the contours shown in (b). The arrows represent (S$_{x}$, S$_{y}$) components of the spin texture. The $S_z$ component is negligible and it is not included.
	}
	\label{fig:3}
\end{figure*}


\textit{Spin-resolved electronic structure.} The electronic properties of bulk WTe$_2$ (P$_{\uparrow}$) calculated from first-principles are summarized in Fig. 2. The band structure along high-symmetry $k$-lines in the Brillouin zone (BZ) shown in Fig. 2(a) agrees perfectly with the previous DFT calculations \cite{tsymbal_scienadv, wu2015temperature, nonlinear_magnetotransport}. WTe$_2$ is a Weyl semimetal with a large spin-orbit coupling (SOC) which combined with the lack of inversion symmetry (IS) causes a noticeable splitting of the bands throughout almost entire BZ. The Fermi surface (FS) visualized in Fig. 2(b) consists of two pairs of hole and electron pockets, the former located between the latter along the $\Gamma-X$ direction. Although it is again in a qualitative agreement with the quantum oscillations (QO) and angle-resolved photoemission spectroscopy (ARPES) \cite{quantumosciallations, fermiarcs, komori}, the exact size and shape of features slightly differs between calculations and experiments, being sensitive to the computational details, such as the choice of the exchange-correlation functional, as well as strain and doping of samples. While the topology of the pockets is responsible for anomalous transport properties reported in WTe$_2$ \cite{nonlinear_magnetotransport}, the symmetry and magnitude of REE will depend also on the spin texture.

Figure 2(c) shows the spin texture of the Fermi surface calculated along the energy isocontours $E=0$ within the $k_x-k_y$ plane for $k_z=0$. Even though at first glance it resembles a Rashba-like pattern, additional radial components manifest as spin winding around each contour. The character of the spin texture can be understood based on the symmetry analysis of the point group $C_{2v}$ that describes the $\Gamma$ point. The mirror symmetry $M_x: (x, y, z) \rightarrow (-x, y, z)$ transforms the spin operator ($S_{x}$, $S_{y}$, $S_{z}$) $\rightarrow$ ($S_{x}$, $-S_{y}$, $-S_{z}$) which is reflected in the left-right symmetry of the spin pattern. Similarly, the mirror symmetry $M_y: (x, y, z) \rightarrow (x, -y, z)$ yields ($S_{x}$, $S_{y}$, $S_{z}$) $\rightarrow$ ($-S_{x}$, $S_{y}$, $-S_{z}$) and manifests as the up-down symmetry. The countours thus do not show standard clockwise or counterclockwise circulation known from the standard Rashba spin-momentum locking. In fact, WTe$_2$ possesses a Rashba-Dresselhaus spin texture, in agreement with the classification presented in Ref. \cite{zunger}. Further properties can be also easily inferred from Fig. 2(c); the pairs of bands $1$,$2$ and $3$,$4$ have opposite circulation as they correspond to spin-orbit splitting of the same band. Moreover, we note that P$_{\uparrow}$ and P$_{\downarrow}$ are connected by the inversion symmetry, thus the signs of the spin texture are opposite (not shown). Overall, the spin-resolved electronic structure of WTe$_2$ is again in a good agreement with the recent experimental study \cite{komori}.

Figure 3 illustrates the electronic structure and spin texture of the bilayer WTe$_2$. The bands along the high-symmetry lines shown in Fig. 3(a) are very similar to the previous work \cite{ren}. The Fermi contours consisting of four bands are illustrated in Fig. 3(b), while a more detailed view in Fig. 3(c) includes also the spin texture along the contours. The spin pattern resembles Rashba spin-momentum locking and again, it can be explained in terms of symmetries. The $\Gamma$ point in this case is described by the point group $C_s$ which implies a mirror symmetry $M_x$ that leaves the $S_x$ component unchanged and reverses the $S_y$ component. This is reflected in the left-right symmetry of the plot. From comparison of Fig. 2(c) and Fig. 3(c), we can see that the effective two-dimensional (2D) Rashba spin-orbit field (SOF), $\Omega_R = (-\alpha_x k_y, \alpha_y k_x)$ dominates in the case of bilayer, whereas in the bulk the Dresselhaus SOF, $\Omega_D = (\beta_x k_x, -\beta_y k_y)$ competes with the Rashba. This is not suprising because Dresselhaus spin-orbit coupling is known to be a bulk property \cite{dresselhaus}.



\begin{figure*}[ht]
		\includegraphics[width=\textwidth]{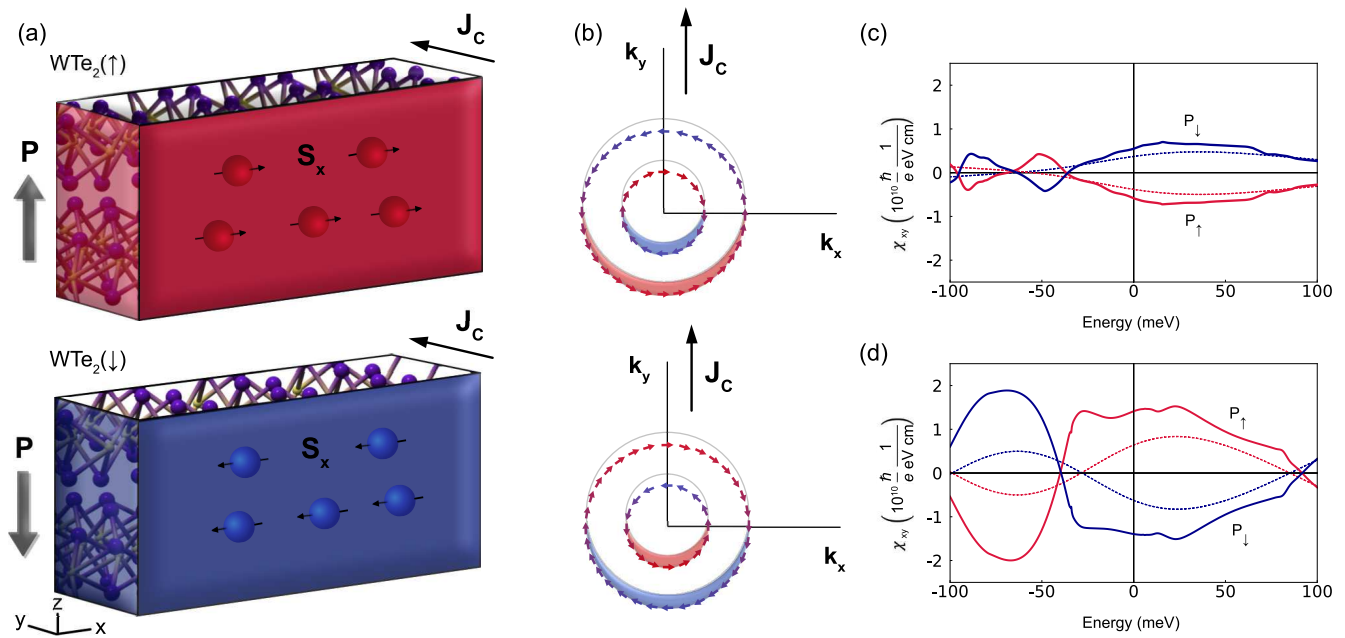}
\caption{Ferroelectric control of REE in WTe$_2$. (a) Schematic illustration of REE in bulk/bilayer WTe$_2$ with electric polarization $P_{\uparrow}$ and $P_{\downarrow}$ shown in upper and bottom panel, respectively. In both cases, the charge current flows along the $y$ direction and the spin accumulation is parallel or anti-parallel to $x$. (b) Illustration of REE in the reciprocal space for a simple Rashba model; the upper and bottom panel represent opposite chirality of spin. When the current flows along $y$, it induces the imbalance of spins along $x$ ($\chi_{xy}$), as marked around the shaded regions which represent the variation of electron distribution. (c) The Edelstein tensor $\chi_{xy}$ vs chemical potential calculated for the bulk according to the geometry shown in (a). The curves corresponding to opposite ferroelectric ground states $P_{\uparrow}$ and $P_{\downarrow}$ are represented as red and blue, respectively. The solid and dashed lines correspond to temperatures $T=0$ and $T=300$ K, respectively. (d) The same as (c) for bilayer WTe$_2$. The volume equivalent was calculated using the effective thickness of WTe$_2$ equal to 15 \AA.}
	\label{fig:4}
\end{figure*}

\textit{Rashba-Edelstein effect.} The principle of the ferroelectric control of the charge-to-spin conversion based on REE is schematically illustrated in Fig. 4(a). The two ground states with opposite ferroelectricity can be switched under an out-of-plane electrical bias, which additionally leads to the sign reversal of spin texture, similarly as in ferroelectric Rashba semiconductors \cite{fersc_nanolett, fersc_adv, fersc_gete}. Because the Rashba-Edelstein effect directly depends on the spin operator \cite{chiral_tellurium}, one can expect that the sign of spin accumulation could be externally manipulated via the changes in electric polarization. Such mechanism is pictorially explained in Fig. 4(b) with the help of a simple Rashba model; different signs of spin texture generate opposite spin accumulation for the same direction of the applied electric current. Even though the spin-resolved electronic structure of WTe$_2$ differs from a standard Rashba system in terms of number and shape of Fermi sheets as well as the winding of the spin texture (see Fig. 2 and Fig. 3), we can show, based on the symmetry analysis and DFT calculations, that the conventional REE will be present both in the bulk and bilayer systems and it would reverse the sign upon the ferroelectric switching.

The Rashba-Edelstein effect will be quantitatively expressed as the spin accumulation calculated in the framework of the semiclassical Boltzmann transport theory. From the linear response, the spin accumulation $\delta \mathbf{S}$ induced by charge current $\mathbf{j}$ can be written as:
\begin{equation}
\delta S^j=\chi^{ji} j_{i}
\end{equation}
The Edelstein tensor $\chi$ can be calculated within the constant relaxation time approximation following the derivation given in Ref. \cite{chiral_tellurium}. The convenient expression implemented in \textsc{paoflow} is the following:
\begin{equation}
  \label{tensor}
  \chi^{ji} = -\cfrac{\sum\limits_{\bm{k}} \langle\bm{S}\rangle_{\bm{k}}^j\bm{v}_{\bm{k}}^i  \frac{\partial f_{\bm{k}}}{\partial E_{\bm{k}}}}{e \sum\limits_{\bm{k}} (\bm{v}_{\bm{k}}^i)^2 \frac{\partial f_{\bm{k}}}{\partial E_{\bm{k}}}}
\end{equation}
where $\mathbf{S}$ stands for the spin operator, $\bm{v}_k$ is the group velocity and $\partial f_{\bm{k}}/\partial E_{\bm{k}}$ denotes the derivative of the Fermi-Dirac function. Similarly to spin Hall conductivity tensor \cite{roy}, different components will yield distinct configurations of electric current and spin accumulation. Based on symmetry arguments, we concluded that the space group of bulk WTe$_2$ (\#31) allows only two independent non-zero components of REE, $\chi_{xy}$ and $\chi_{yx}$. Even though the reduced symmetry of the bilayer (\#6) additionally permits for $\chi_{zy}$ and $\chi_{yz}$ elements, they will be zero because the $S_z$ spin texture is hardly present and the electric current $j_z$ cannot flow across a 2D system.

Figure 4(c) shows the magnitude of $\chi_{xy}$ as a function of the chemical potential in bulk WTe$_2$ for both ferroelectric configurations $P_{\uparrow}$ and $P_{\downarrow}$. These states manifest exactly opposite response which demonstrates that CSC could be manipulated by changes of the electric polarization. While the entire energy window shown in the plot could be reached via gating, the maximum value of the spin accumulation lies close to the Fermi level. The magnitude at $E_F$, $\chi_{xy} (E_F)$ = 0.55 $\times 10^{10} \frac{\hbar}{e}\frac{1}{\mathrm{eV cm}}$ means that a charge current of 100 A/cm$^2$, will generate the spin accumulation of $0.55 \times 10^{12}/$cm$^{3}$. Although the values of REE reported for different materials are very scarce, we can still compare the result with the previously calculated spin accumulation for Te ($\sim 10^{13}/$cm$^{3}$) and the Weyl semimetal TaSi$_2$ ($0.4\times 10^{12}/$cm$^{3}$) \cite{chiral_tellurium}. It is evident that CSC in WTe$_2$ resembles more the latter, which can be attributed to the presence of states with opposite spin polarization around the Fermi level, that generate the opposite contribution to the spin accumulation.

In Fig. 4(d), we report the REE calculated for the ferroelectric bilayer. The magnitudes are larger than in the case of bulk, for example $\chi_{xy} (E_F)$ = 1.4 $\times 10^{10} \frac{\hbar}{e}\frac{1}{\mathrm{eV cm}}$, suggesting that the CSC could be twice stronger in a thin film. The difference can be understood by comparing spin patterns along the constant energy contours for bulk and bilayer shown in Fig. 2(c) and Fig. 3(c), respectively. In the bulk WTe$_2$, the contribution from the Rashba ($\propto \alpha_x$) and Dresselhaus SOF ($\propto \beta_x$) compete, which results in a lower value of the component $\chi_{xy}$. In contrast, in the bilayer the relative strength of Rashba SOF ($\propto \alpha_x$) is larger, inducing a greater response for the same tensor element. We also note that the small SOC splitting in the bulk results in nearly degenerate FS pockets (3 and 4, see Fig. 2c) and their contributions will nearly cancel each other, yielding lower REE, while the splitting of the FS in the bilayer WTe$_2$ is stronger. The values of $\chi_{yx}$ are overall slightly lower than $\chi_{xy}$, and are summarized for completeness in the Supplementary Material (SM) \footnote{See Supplemental Material at [http://link.aps.org/supplemental/
	10.1103/PhysRevMaterials.6.L091404] for the plots of the Edelstein tensor calculated at different temperatures.}.

Last, we will discuss the perspectives of applications of WTe$_2$ in spintronics devices. One of the essential aspects is the possibility of operating at room temperature. We note that the spin-orbit splitting of the bands close to the Fermi level is large (well above 25 meV) which suggests that the spin texture would be robust against thermal effects. Moreover, we roughly estimated REE for finite temperatures. The results for $T=300$ K are shown in Fig. 4(c) and 4(d) (dashed lines), while the additional data for $T=150$ K are reported in the SM. Even though the magnitudes drop at room temperature, the REE would decrease by only a factor of 2 and it could be still detected. The second aspect relevant for ferroelectric spintronics is the robustness and switching of the electric polarization. To the best of our knowledge, WTe$_2$ is the only confirmed ferroelectric metal at room temperature, providing a unique opportunity of ferroelectric control of (spin) transport phenomena. However, the small values of polarization might be an obstacle for devices, especially those based on the bilayer WTe$_2$ whose polarization is significantly lower than in conventional ferroelectrics; thus the usefulness for electronics must be ultimately verified by experiments.


\textit{Conclusion.} In summary, we have predicted for the first time the magnitudes of the Rashba-Edelstein effect in ferroelectric WTe$_2$ at different temperatures. We confirmed that (i) the current-induced spin accumulation switches the sign upon the reversal of the electric polarization, (ii) the CSC can be realized at room temperature, (iii) REE is present also in the bulk despite the fact that the spin texture does not have the conventional Rashba spin-momentum locking. While the role of rather small electric polarization needs to be further examined, WTe$_2$ has potential for variety of applications in electronic devices. In particular, the unusual bulk REE could be utilized for spin-transfer torques, while the bilayer seems promising for ferroelectric spin-orbit (FESO) devices, i.e. the analogs of MESO in which the magnetoelectric material is replaced by a ferroelectric with strong CSC. We believe that these results will stimulate experimental studies of ferroelectric control of spin transport in WTe$_2$ that can be highly efficient due to the co-existence of ferroelectricity and semimetallicity.

\begin{acknowledgments}
We thank Maxim Mostovoy and Marcos Guimar\~aes for helpful discussions. J.S. acknowledges the Rosalind Franklin Fellowship from the University of Groningen. The calculations were carried out on the Dutch national e-infrastructure with the support of SURF Cooperative (EINF-2070), and on the Peregrine high performance computing cluster of the University of Groningen.
\end{acknowledgments}

\end{document}